\documentclass{article}
\usepackage{arxiv}
\usepackage[utf8]{inputenc} 
\usepackage[T1]{fontenc}    
\usepackage{hyperref}       
\usepackage{url}            
\usepackage{booktabs}       
\usepackage{amsfonts}       
\usepackage{nicefrac}       
\usepackage{microtype}      
\usepackage{lipsum}
\usepackage[dvipsnames]{xcolor}
\usepackage{graphicx}
\usepackage[natbibapa,nodoi]{apacite}

\title{Astronomers' and Physicists' Attitudes Toward Education \& Public Outreach: A Programmatic Study of The Dark Energy Survey}

\author{Farahi,~A.\textsuperscript{1}, Gupta, R.~R.\textsuperscript{2}, Kraweic,~C.\textsuperscript{3}, Plazas,~A.~A.\textsuperscript{4}, Wolf,~R.~C.\textsuperscript{5}\\\textsuperscript{1}{Department of Physics, University of Michigan}\\\textsuperscript{2}{Physics Division, Lawrence Berkeley National Laboratory}\\\textsuperscript{3}{Department of Physics and Astronomy, University of Pennsylvania}\\ \textsuperscript{4}{Department of Astrophysical Sciences, Princeton University}\\ \textsuperscript{5}{AAALab, Graduate School of Education, Stanford University}\\\\All authors contributed equally to the work presented in this manuscript.\\CONTACT: \texttt{rcwolf@stanford.edu}\\\\A version of this manuscript was published in \textit{Journal of STEM Outreach}, June 2019.}

\begin{document}

\maketitle

\begin{abstract}
We present a programmatic study of physicists’ and astronomers’ attitudes toward education and public outreach (EPO) using 131 survey responses from members of the Dark Energy Survey. We find a disparity between the types of EPO activities researchers deem valuable and those in which they participate. Most respondents are motivated to engage in EPO by a desire to educate the public. Barriers to engagement include career- and skill-related concerns, but lack of time is the main deterrent. We explore the value of centralized EPO efforts and conclude with a list of recommendations for increasing researchers’ engagement.
\end{abstract}


\section{Introduction}
\label{sec:Intro}

Over the past twenty years, the need for improved communication between scientists and the general public has been recognized worldwide \citep{Burns:2003, Kenney:2016, NRC:2010}. Advances in science and technology have transformed life in the 21st century, and institutions ranging from government agencies to business conglomerates are calling for change in the perception and understanding of science. Such a paradigm shift has been discussed in the context of the science, technology, engineering, and mathematics (STEM) disciplines, eliciting reform in education materials spanning from the classroom to informal education spaces. The implementation of the Broader Impacts criterion by the National Science Foundation (NSF) in 1997 set a major milestone in reaching the goals of education and public outreach (EPO) among U.S. researchers and educators. This pivotal charge requires scientists and educators who are seeking federal funding to propose formal and informal EPO programs. The current framework encourages research proposals which aim to broaden participation of under-represented groups, develop enhanced infrastructure for novel forms of research and education, enhance the scientific and technological perception of the public, or provide general benefit to society.

Today, there are striking differences between the views of the public and scientists on many science-related topics \citep{Pew:2015,Stocklmayer:2012}. Additionally, the public does not know much about the scientific process or academic culture, nor do the scientists know much about the public interest \citep{Levy:1992, Miller:1998}. Therefore, scientific societies such the American Association for the Advancement of Science are advocating a model in which scientists engage with the public in meaningful dialogue that positively impacts the attitudes and behaviors of not only the general public, but of the scientists themselves.\footnote{\url{https://mcmprodaaas.s3.amazonaws. com/s3fs-public/content_files/2016-09-15_AAAS-Logic-Model-for-Public-Engagement_Final.pdf}}

Of the many STEM fields, astronomy is one of the most popularly used to spur public and academic interest \citep{Heck:2013}. The night sky is accessible across the globe and provides a spark for curiosity. Astronomical images can be both scientifically discussed and aesthetically admired. Questions surrounding the origin and fate of the Universe inspire scientific, moral, and philosophical debate. Given the natural curiosity inspired by the subject, it is no surprise that there is overwhelming evidence that the public is interested in astronomy programming. Each year nearly 28 million people visit planetaria \citep{NRC:2001}, and hundreds of thousands make their way to astronomical observatories. On social media, the NASA Twitter account has roughly 30 million followers to date.\footnote{\url{https://twitter.com/nasa}} Interest in the academic discipline of astronomy has also increased significantly over the past thirty years. Since 1983, the number of universities in the U.S. offering graduate degrees in astronomy has grown by over 50\%, and the number of students receiving bachelor’s degrees has nearly doubled \citep{Mulvey:2014}. Enrollment in introductory astronomy courses is approaching nearly 200,000 students, an increase of roughly 10\% in the last decade. 
 
 It would seem that such a demand for astronomy material would encourage the larger community of astronomy professionals (including physicists, astrophysicists, astronomers, telescope engineers, and technical support staff) to engage in both formal and informal EPO. In the cases of professionals employed at universities, engagement in formal astronomy and/or physics education is often mandated. However, as in many other STEM disciplines, there remains a disconnect between the duties of the professional and engagement in other types of EPO. A recent study of researchers’ (in higher education, research institutes, and clinical settings) engagement in the United Kingdom concluded “that public engagement is more firmly embedded in the context of the arts, humanities and social sciences than it is among researchers in science, technology, engineering and mathematics” \citep{Hamlyn:2015}.  This is particularly evident in the STEM fields as the “Sagan Effect,” a perceived stigma imposed by colleagues in academia on those research professionals who are actively involved in public engagement \citep{Shermer:2002}. In an analysis of 97 interviews with academic biologists and physicists, \citet{Ecklund:2012} found that their respondents believed that a change in attitude toward science outreach is difficult to achieve. In a survey of 59 physicists, \citet{Johnson:2014} observed that public engagement is considered to be outside the realm of professional tasks and that those who participate in outreach activities are “perceived as occupying a marginal status.”
 
Scientists’ views of and engagement  in EPO has recently become a topic of research. With the increasing number of public funding agencies mandating “broader impact” components in research grants, \citet{Andrews:2005} surveyed and interviewed scientists about their participation in and motivations for public outreach as well as impediments to involvement. In one of the first systematic surveys of a large international group of astronomers, conducted during the 2012 General Assembly of the International Astronomical Union, \citet{Dang:2015} observed that 79\% of respondents ($n = 155$; where $n$ denotes the total number of responses) expressed belief that EPO initiatives are essential. In addition, only 43\% ($n = 116$) of a subsample of respondents (several respondents elected not to answer survey items concerning funding) were explicitly funded to engage in EPO programming. \citet{Dang:2015} also asked about barriers to EPO engagement, finding that lack of time and grant funding were significant deterrents. Such barriers were confirmed by survey responses and interviews by \citet{Johnson:2014} and \citet{Thorley:2016}. Similarly, \citet{Ecklund:2012} found that while 58\% of their respondents were involved in science outreach, they nevertheless cited barriers to engagement coming from their institutions, the public, and even scientists themselves. A survey of academics from multiple STEM and humanities disciplines offers a different perspective: how researchers perceive their colleagues’ attitudes toward and participation in EPO is a better predictor of engagement than funding or time constraints \citep{Poliakoff:2007}. Clearly, as \citet{Johnson:2014} assert, better ``understanding how scientists interpret outreach'' is crucial for both research and policy. Furthermore, understanding these scientists’ perspectives of EPO will be essential for professionals developing future astronomy-related EPO programs.

In this article we present a programmatic study of researchers’ EPO experience as compiled from 131 survey responses from physicists, astronomers, and astrophysicists affiliated with the Dark Energy Survey\footnote{\url{www.darkenergysurvey.org}} \citep[DES; ][]{DES}. We believe this study is unique in both scope and perspective in that at the time the survey was administered, the authors were all practicing astrophysicists themselves and early-career members of the DES collaboration. In the context of this article, we define EPO in as any type of engagement between a STEM professional and a non-expert, including, but certainly not limited to: K-12 curriculum development and classroom visits, undergraduate teaching and mentorship, science festivals, written communication, social media, public lectures, radio and TV appearances, and museum programming.  DES is an international collaboration of hundreds of researchers primarily working together to study the effects of dark energy. The collaboration is composed mainly of faculty, staff scientists, postdoctoral researchers (post-docs), and graduate students. It is structured into several smaller groups of researchers studying a particular scientific focus, where members may overlap between areas of interest. Hereafter we refer to these collections of researchers as ``working groups.'' Since its inception in 2014, the DES Education and Public Outreach Committee has acted as a working group, developing and cultivating a diverse repertoire of online and in-person EPO initiatives. A brief description of the DES organizational structure and the DES EPO program is provided in Section~\ref{sec:DES}.

Throughout this article, we explore scientists’ engagement with EPO from multiple perspectives, including their motivations and barriers, the types of activities in which they engage, and the types of activities which they value. Findings are presented in Section~\ref{sec:Results}, and interpretations of these findings are discussed in Section~\ref{sec:summary}. We consider both general EPO engagement and involvement specific to the DES EPO program, and note that while we refer to the collective group of DES members as ``researchers,'' the attitudes expressed do not reflect the opinions of all professional scientists or researchers. The survey and data analyses presented here are intended to not only contribute to a better understanding of scientists’ relationship with EPO, but also to inspire topics of future, more targeted research that may be of interest for researchers and EPO facilitators alike. Opportunities for such research are described in Section~\ref{sec:limitations}. 

\section{The Dark Energy Survey Collaboration}
\label{sec:DES}

\subsection{Project Structure}
DES traces its origins as a project concept back to at least 2004, and the first DES images were taken in September 2012. Support for DES is provided by grants primarily from the U.S. Department of Energy and the National Science Foundation. The larger DES collaboration is comprised of researchers from universities, research institutes, and astronomical observatories around the globe. 

The internal organization of DES is divided into three main components: collaboration affairs (i.e., managers of collaboration-wide tasks, including the DES director), science, and operations. Each of these three committees is further subdivided into a variety of smaller groups, e.g., the Science Committee is comprised of the science working groups. We note that EPO was not a foundational part of the DES organizational structure. Further details about DES science and organizational infrastructure is discussed in \citet{Wolf:2018}. 	

\subsection{The DES Education and Public Outreach Committee}
\label{subsec:EPOC}
The DES Education \& Public Outreach Committee (EPOC) became a part of the official DES organizational infrastructure in the fall of 2014. The evolution of the EPOC was unique at this scale in astronomy, as it was a grass-roots effort initiated, managed, and implemented by practicing researchers. At the time, the three founders of the EPOC were active members of DES: an assistant professor based in the United Kingdom and a post-doc and graduate student based in the United States. The roles and responsibilities of the EPOC evolved organically, as there were no policies in place for how the EPOC and its programming should interact and coordinate with the rest of the collaboration. As the primary organizers of EPO for the collaboration, the EPOC oversaw and contributed to: maintenance of the DES website, DES social media, creation of informal and formal educational materials, DES events with local communities (e.g., museum events and science fairs), internal EPO reporting, public relations, and more. The centralized DES EPO program had a limited, floating budget per the discretion of the DES director, which was jointly funded by the DES collaborating institutions. Further discussion of DES EPO programming is outlined in \citet{Wolf:2018}. 

During the first year of the EPOC, most programming was organized and executed by the EPOC founders. The EPOC prioritized updating, enhancing and maintaining the DES online presence, largely because collaboration members could contribute to online EPO initiatives from anywhere in the world. In addition to the weekly coordinator meetings, monthly EPOC teleconferences with collaboration members interested in participating in EPO were established to discuss the progress of the various projects. One of those projects included internal communication and resulted in a monthly DES-EPO newsletter that was sent electronically to every registered email in the DES membership database (hereafter the DES LISTSERV). 

To reach as many collaboration members as possible, the EPOC founders advertised EPO activities (at collaboration meetings, through collaboration-wide emails, in the monthly newsletter, etc.) as opportunities for formal engagement with students and informal engagement with the general public. The EPOC hypothesized that establishing an inclusive understanding of EPO which encompassed the gamut of activities would appeal to the vast range of experience and positions in the collaboration membership. 

\section{Measuring Dispositions Toward EPO}
\label{sec:Meas}

\subsection{Motivation}
\label{subsec:motivation}

One of the most significant challenges of the DES EPOC was sustaining projects which relied on collaboration-member participation. This issue was two-fold: it was both difficult to inspire new interest and participation and difficult to maintain active involvement once programs were initiated. As EPO is now becoming integrated into infrastructure for new projects, e.g., the Large Synoptic Survey Telescope,\footnote{\url{https://www.lsst. org/about/epo}} we sought to use DES EPO as a programmatic study to explore how participation in EPO can be sustained as a model for future science collaborations. This led to the following exploratory Research Questions:
\begin{enumerate}
    \item How are DES members spending their time engaging in EPO? 
    \item What motivates a DES member to participate in EPO activities?
    \item What deters a DES member from participating in EPO activities?
    \item How can collaborations and organizations best support member participation in EPO activities?
\end{enumerate}

We hypothesized that responses to these research questions would vary by researcher demographics such as age and employment position. For example, we hypothesized that researchers with temporary  positions (e.g., graduate students and post-docs) may spend less time on EPO activities than those with permanent positions as they may feel compelled to spend more or all of their time on research in pursuit of a permanent position. Where possible, differences in EPO dispositions by demography are discussed in Section~\ref{sec:Results}.

\subsection{Survey Design}
\label{subsec:survey}

To investigate these Research Questions, the EPOC founders designed a survey to be distributed collaboration-wide. The survey was intended to cover a diverse array of topics related to the Research Questions, while also maintaining a reasonable length to encourage participation. Language in the survey was chosen to mimic the ways in which EPO had been discussed throughout collaboration events. To provide respondents with a guiding mental framework, the introduction of the survey defined STEM EPO under the umbrella of the \citet{Burns:2003} ``vowel analogy'' of science communication. Specifically, the survey stated the following:

\fbox{\begin{minipage}{\textwidth}
For the purpose of this survey, we will define STEM EPO as all activities which fall under the umbrella of ``science communication'' as defined in Burns et al. 2012. This includes undergraduate teaching, participation in science festivals, mentoring, etc.\\

SCIENCE COMMUNICATION (SciCom) may be defined as the use of appropriate skills, media, activities, and dialogue to produce one or more of the following personal responses to science (the vowel analogy) \\

\hspace{1cm}\textbf{A}wareness, including familiarity with new aspects of science \\

\hspace{1cm}\textbf{E}njoyment or other affective responses, e.g. appreciating science as entertainment or art \\

\hspace{1cm}\textbf{I}nterest, as evidenced by voluntary involvement with science or its communication \\

\hspace{1cm}\textbf{O}pinions, the forming, reforming, or confirming of science-related attitudes \\

\hspace{1cm}\textbf{U}nderstanding of science, its content, processes, and social factors \\

Science communication may involve science practitioners, mediators, and other members of the general public, either peer-to-peer or between groups.
\end{minipage}}

Survey language was discussed within the EPOC and with professional science communicators (e.g., laboratory communications officers). We note that the published version of the survey mistakenly quotes the Burns article year as 2012. This was clarified for collaboration members who asked for more information.

Although survey wording went through multiple iterations, the survey was not externally tested or validated via focus groups before it was administered. Examination of survey responses illuminated points of confusion (e.g., unclear language) or incomplete item response options (e.g., a multiple choice item did not include a popular “Other” option). These are discussed as limitations of the analysis and opportunities for future work in Section~\ref{sec:limitations}.

The survey was created using the Google Forms platform.\footnote{\url{https://www.google. com/forms/about/}} It was electronically disseminated to collaboration members and conserved respondent anonymity. Specific directives were provided at the beginning of each section of the survey to frame the definition of EPO relevant to that section. The survey was composed of three sections: 1) an introduction, 2) questions about general STEM EPO engagement, and 3) questions about EPO attitudes specific to the structure of and resources available to DES and other large science collaborations. A final section collecting demographic information (i.e., gender, ethnicity, age, and position) concluded the survey. While all questions in the demographic section of the survey were mandatory, each question provided respondents with the option to decline a response. The survey consisted of mixed question types including Likert measures (scaling method used to gauge response to a statement, i.e., the extent to which a respondent agrees or disagrees), multiple choice and checkbox questions, and free response. We note that due to nuances with the survey platform, in some cases respondents could not change an incorrectly submitted response. The survey was open to participants for two weeks; reminder emails were sent with one week, three days, and one day remaining in the open survey period.

\begin{figure}[htp]
    \centering
    \includegraphics[width=\textwidth]{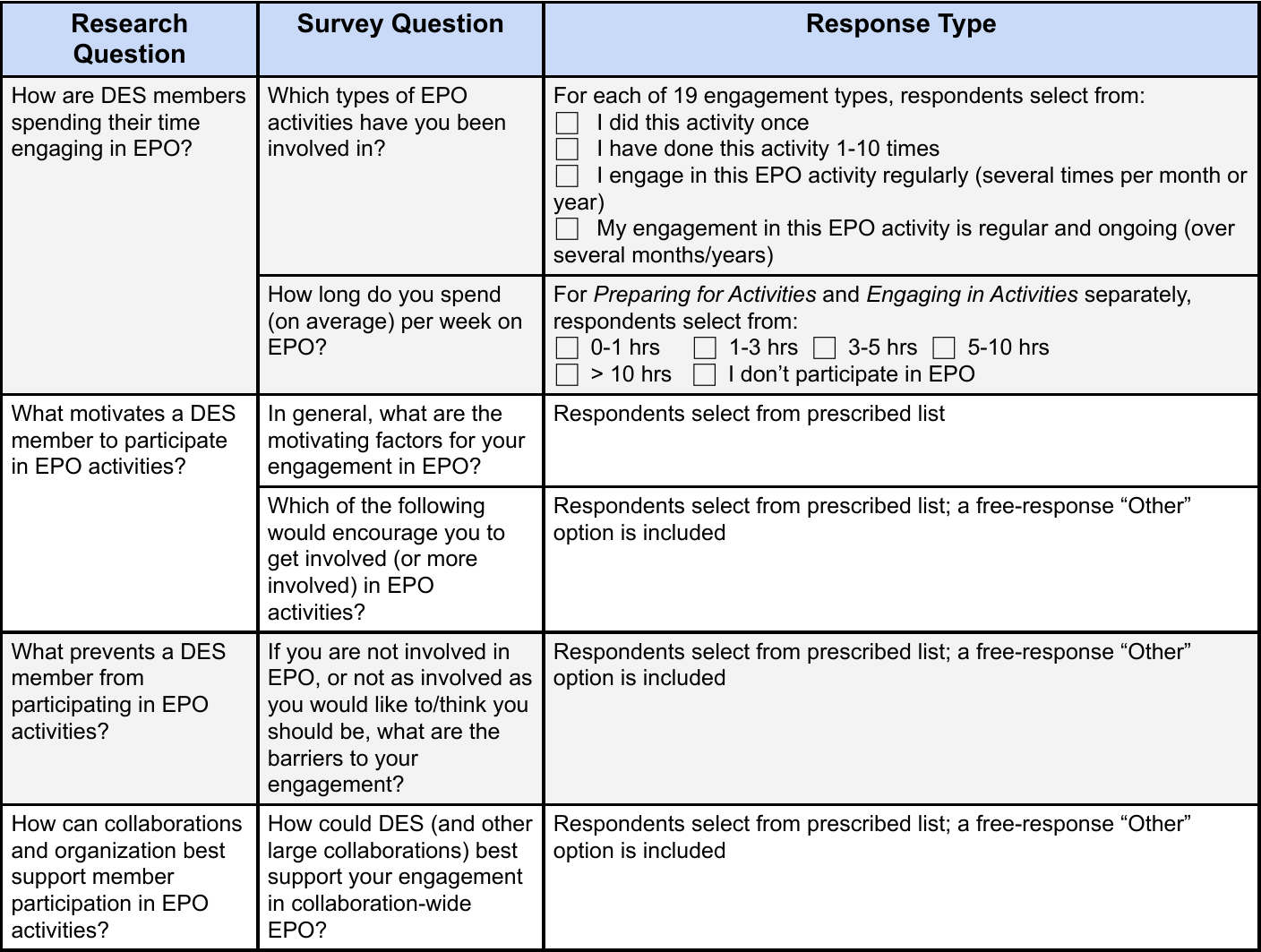}
    \caption{Example survey items. The full survey is provided in the supplemental materials.}
    \label{fig:ExTab}
\end{figure}

We investigated researchers’ dispositions from multiple perspectives by including thirty survey items related to diverse components of the EPO experience. Respondents were asked about the types of activities in which they have engaged and how frequently that engagement takes place. We inquired about personal and professional motives for engagement, as well as any barriers. Furthermore, we asked researchers to describe how their peers view EPO and to provide their feedback on more centralized EPO organizational efforts. Example survey items are provided in Figure~\ref{fig:ExTab}. The complete survey and (de-identified) data are provided as supplementary materials to the article and on the DES EPO research website.\footnote{\url{https://www.darkenergysurvey.org/education/des-education-outreach-science-communication-research/}}

\subsection{Respondents}
\label{subsec:Respondents}
All DES members are encouraged to subscribe to the DES-wide LISTSERV, which is frequently used for collaboration-wide announcements and updates. The survey described in the previous section was emailed to the DES LISTSERV, which at the time of this study, included 606 subscribers. Subscribers include current active DES members, as well those who are either inactive or have since left the field.

\begin{figure}[htp]
    \centering
    \includegraphics[width=\textwidth]{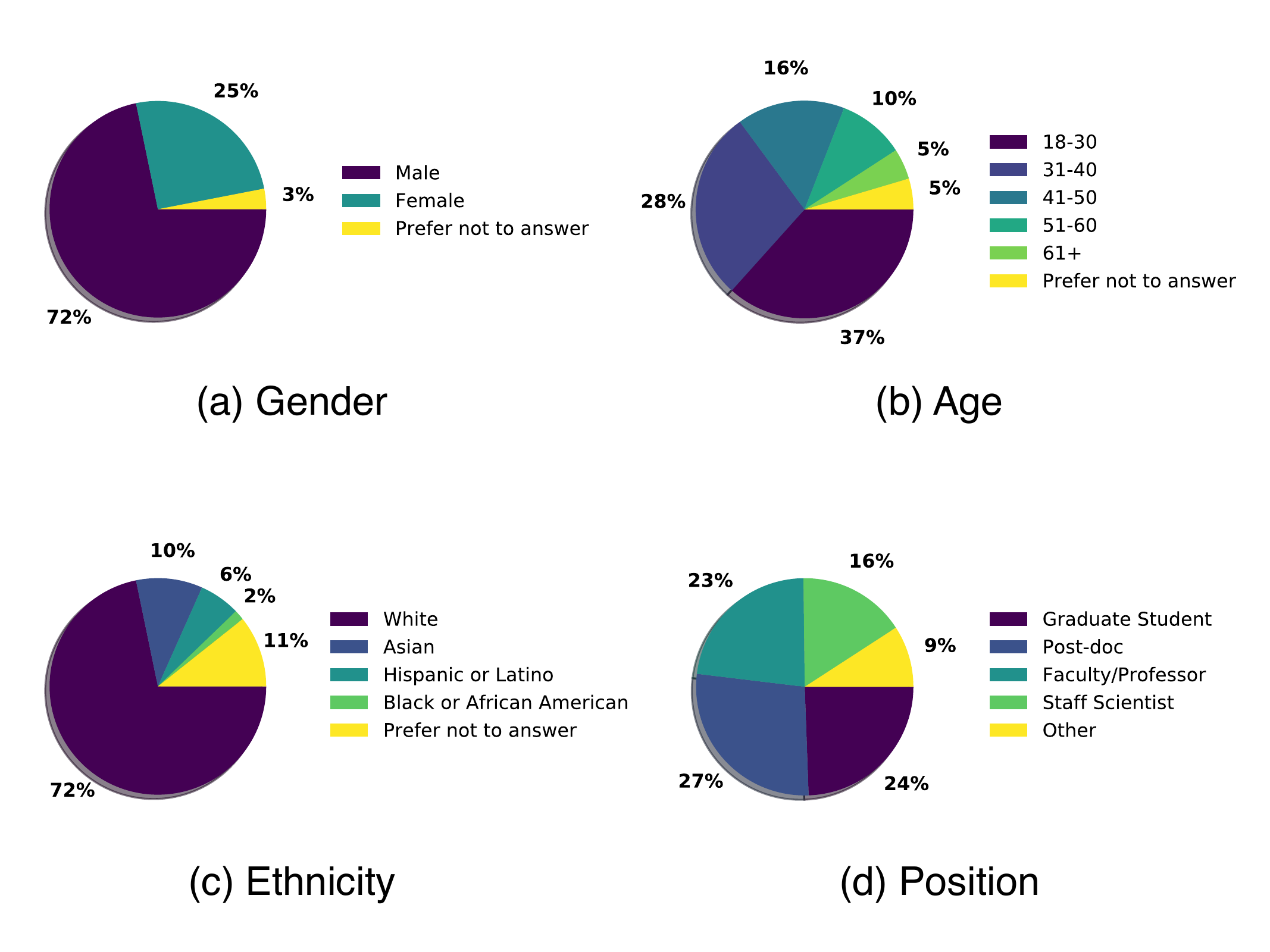}
    \caption{Demographic information for study respondents. Here we display breakdowns by (a) gender, (b) age, (c) ethnicity, and (d) current position as self-reported in the survey.}
    \label{fig:Fig1}
\end{figure}

In total, 131 current and former DES members (22\% of the LISTSERV membership) participated in the online survey, of which 115 self-identified as ``Active Members.'' Figure~\ref{fig:Fig1} displays distributions of respondent gender, age, ethnicity, and employment position. Respondents were predominantly male and white. Most were relatively early career scientists: 65\% reported they were under the age of 40 and 37\% were younger than 30. Respondents were more evenly distributed with respect to current position. Post-docs, graduate students, and faculty each composed roughly a quarter of those surveyed. The remaining quarter consisted of staff scientists and people with other occupations (such as science educator, scientist emeritus, and software developer). These correspond to Questions 25-29 in the survey; for clarity we will refer to specific survey questions by number throughout (e.g., Q25-29).

The DES membership database records are not current or detailed enough to allow us to make demographic comparisons of the respondents to the full DES collaboration. It is, however, possible to compare to recent data drawn from the larger astronomy community, such as the American Astronomical Society (AAS) Workforce Survey of 2016 US Members.\footnote{\url{https://aas.org/ files/aas_members_workforce_survey.pdf}} The AAS survey results consist of responses from 1795 AAS members living in the U.S. Of the AAS respondents, 73\% identified as male compared to 72\% in our survey. One percent of AAS respondents and 3\% of DES respondents preferred not to indicate their gender. The distribution of ethnicities for AAS respondents was 84\% white, 9\% Asian, 3.5\% Hispanic or Latino, and 1\% black or African American. The corresponding fractions for our DES respondents were 72\%, 10\%, 5\% and 1.5\%. Four percent of AAS respondents and 11\% of DES respondents preferred not to indicate their ethnicity. With respect to employment type, 17\% of AAS respondents were temporary while the remainder were ``Potentially permanent.'' By contrast, the majority of our DES respondents were temporary (51\% were graduate students or post-docs). We conclude that our survey appears to sample similar demographics to those captured in the recent AAS survey of US members in terms of gender and ethnicity, but that our respondents contain a much higher fraction of people with temporary employment. To explore questions for which we had hypotheses about response differences between demographic groups (and for which there was sufficient sample size), respondents were separated into several categories: gender [male, female], ethnicity [white, non-white], age [18-30, 31-40, 41+], and position [permanent (i.e., Staff Scientist, Faculty/Professor, or Scientist Emeritus), temporary (i.e., Undergraduate Student, Graduate Student, or Post-Doc)]. We note that since age and position are strongly correlated, is it difficult to disentangle which of them is the dominant effect in any observed correlations.

As participation in the survey was entirely voluntary, we could not ensure that all DES members responded or that those who did were a representative sample of the full DES collaboration. Therefore, selection bias is a factor that impacts the results presented here. It is likely that many of the study respondents were members who already had some interest in EPO. Roughly 79\% of respondents stated that they were involved in some type of EPO project local to their institution or community (Q11), and 66\% responded that they had participated in a DES-specific EPO initiative (Q21).

\section{Results}
\label{sec:Results}
In the following sections, we present the primary results of the EPO attitudes survey. We remind the reader that these results are drawn from a select sample of astronomy and physics researchers whose places of work primarily include universities and national laboratories. Many of our respondents are also likely invested in EPO, either as mandated by a university (i.e., undergraduate teaching) or as a hobby or volunteer effort (i.e., giving public lectures, volunteering at museums). As such, we emphasize that the results presented here may not apply to the population of astronomy researchers, and leave a similar study exploring EPO attitudes for a more diverse population (i.e., researchers based at institutions of informal learning) for future work.

\subsection{Types of Engagement and Time Commitment to EPO}
\label{subsec:eng}

\begin{figure}[htp]
    \centering
    \includegraphics[width=\textwidth]{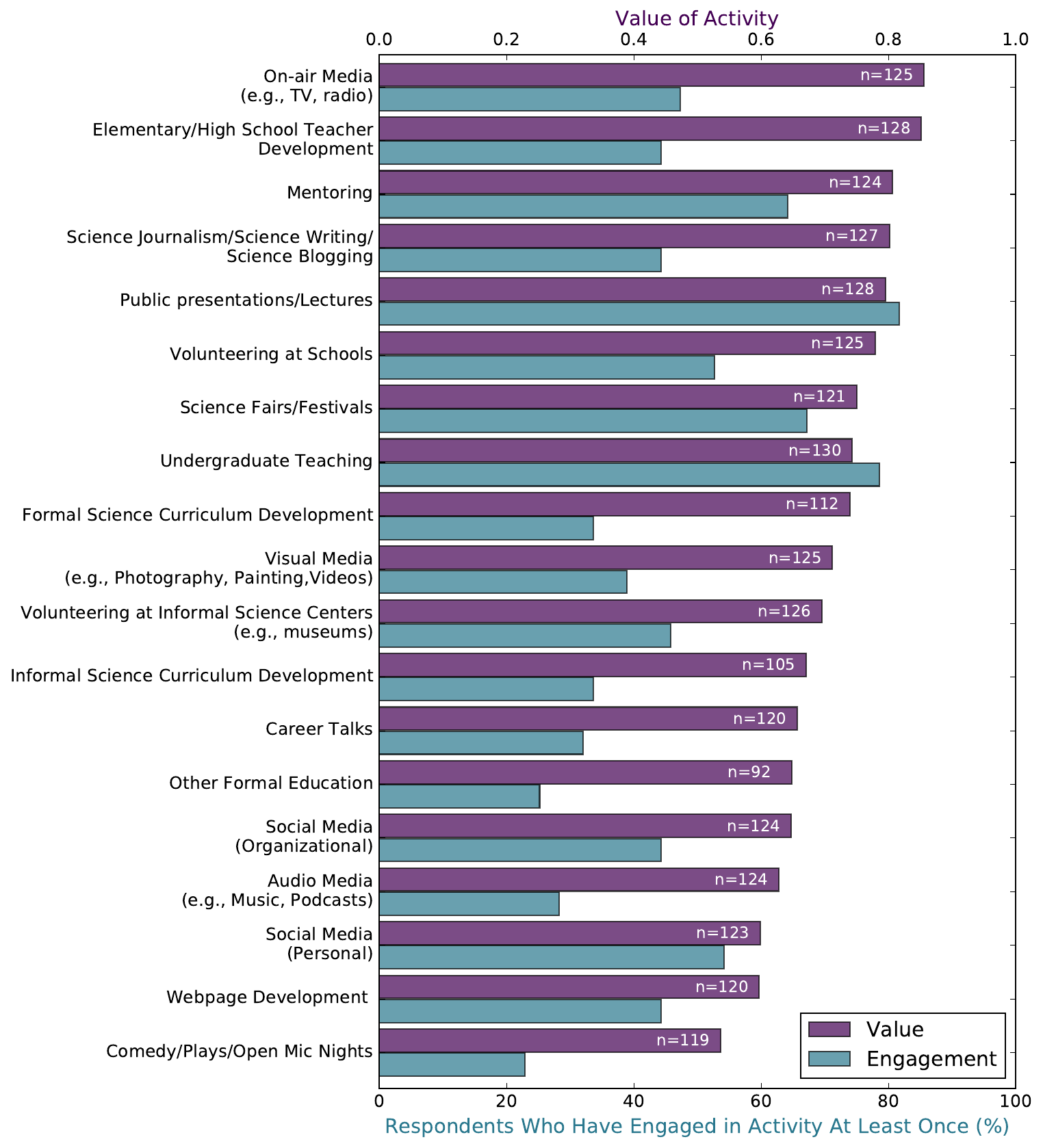}
    \caption{Respondents were presented with 19 activities and asked 1) if they have ever engaged in the activity and 2) to rank its value (level of impact). ``Value of Activity'' is calculated from Eq. 1. The number of counted responses, $n$, is noted in white (out of a total 131 respondents).}
    \label{fig:Fig2}
\end{figure}

Motivated by Research Question 1, the first main section of the survey explored the types of activities in which researchers participate and the time they devote to doing so. We provided a list of 19 EPO activities (see Figure~\ref{fig:Fig2}), spanning a range of engagement audiences, environments, and media, and asked researchers to indicate how frequently (if at all) they had engaged in each (Q7). The five most frequent responses were: \textsc{Public presentations/lectures} (82\%), \textsc{Undergraduate Teaching} (79\%), \textsc{Science fairs/festivals} (67\%), \textsc{Mentoring} (64\%), and \textsc{Social media (Personal, i.e., from a personal Twitter account)} (54\%). We note that for this particular survey item, choosing a response indicates that the respondent had participated in this activity at least once. We find these most common answers unsurprising, as participation in these activities is accessible to, and commonly asked of, scientists at many academic institutions. In fact, \textsc{Undergraduate Teaching} and \textsc{Mentoring} had the highest number of respondents who marked \textsc{My engagement in this EPO activity is regular and ongoing (over several months/years)} at $n$ = 31 and $n$ = 29 respectively. 

Participation frequency in specific EPO activities alone, however, may not be indicative of the importance researchers place upon them. One could posit that pragmatic and logistical factors such as ease, cost, and required time likely influence how researchers elect to engage in EPO. Furthermore, it is possible that these factors are more influential than the perceived value of the activities themselves. To explore this hypothesis, we asked survey respondents to rank the nineteen EPO activities on a 5-point Likert scale: 1 = \textsc{Least Impactful/Valuable to the Audience} to 5 = \textsc{Most Impactful/Valuable to the Audience} (Q6). They were also given the option to choose \textsc{Not impactful/Should not count as EPO} or \textsc{I Don’t Know}. To determine which activities respondents deemed the most impactful (highest value), responses were scored using the following metric for each activity:

\begin{equation}
\label{eq:eq1}
    Value = \frac{R}{5n},\, \mathrm{where}\, R = \sum_{i=1}^{n}R_i \, \mathrm{and} \, n = n_\mathrm{resp} - n_\mathrm{IDK}
\end{equation}

Here $R_i$ is the rank from 0-5 (\textsc{Not Impactful} responses were counted as 0), $n_\mathrm{resp}$ is the total number of responses, and $n_\mathrm{IDK}$ is the number of \textsc{I Don’t Know} responses (which are excluded from the sum). Using Eq.~\ref{eq:eq1}, we find the top five activities with the highest value are: \textsc{On-air media (e.g., TV, radio)}, $Value = 0.86, n = 125$; \textsc{Elementary/High School Teacher Development}, $Value = 0.85, n = 128$; \textsc{Mentoring}, $Value = 0.81, n = 124$; \textsc{Science journalism/science writing/science blogging}, $Value = 0.80, n = 127$; \textsc{Public presentations/lectures}, $Value = 0.80, n = 128$. Figure~\ref{fig:Fig2} directly compares reported participation and perceived value for all 19 activities. Among the largest disparities found in this comparison are  \textsc{Elementary/High School Teacher Development}, \textsc{Science journalism/science writing/science blogging}, and \textsc{On-air media (e.g., TV, radio)}, which are ranked high in value, but are not as commonly engaged in as the other highly-valued activities. Twenty-eight respondents out of the 58 that noted engagement in \textsc{Elementary/High School Teacher Development} checked the \textsc{I did this activity once}
box. Among the activities with the least participation are \textsc{Audio Media (e.g., Music, Podcasts)} and \textsc{Comedy/Plays/Open Mic Nights} which are also among the lowest-valued. After \textsc{Elementary/High School Teacher Development}, \textsc{Audio Media (e.g., Music, Podcasts)} and \textsc{Visual media (e.g., photography, painting, sculpture, animations, videos)}, were the next most commonly noted as activities that participants only engaged in once ($n = 25$ and $n = 24$). \textsc{Social Media (Personal, i.e. from a personal Twitter account)}, rated third lowest in value, was the activity with the most respondents ($n = 19$) indicating \textsc{I engage in this EPO activity regularly (several times per month or year)}. 

We also included several questions designed to probe how much time DES researchers commit to EPO. We asked survey respondents to indicate their average weekly time commitment to preparing and engaging in EPO activities by checking corresponding boxes (Q9). In addition, we asked how much time they \textit{would like} to spend on such tasks (Q10). A summary of these responses is shown in Table~\ref{tab:time}. We find that 88\% of respondents spend three hours or less per week engaging in or preparing for EPO activities. The amount of time respondents would like to spend on preparation (or engagement) is strongly and significantly correlated with the time they currently devote ($r  = 0.99, p < 0.001$).  Thirty-five percent of respondents would like more time to prepare for EPO activities, while 62\% are satisfied with their current preparation time. Similarly, 45\% of respondents would like to spend more time actually engaging in EPO, while 54\% are content with current engagement. Of those respondents who would like to increase their engagement time, 81\% indicated that they would like to increase their engagement by one time interval (e.g., if a respondent currently engages 1-3 hours per week, she would like to spend 3-5 hours per week).  Furthermore, while nearly 10\% of respondents do not engage in EPO, only 3\% lack the interest.

{\renewcommand{\arraystretch}{1.4}
\begin{table}[t]
\caption{Current (a) and desired (b) time commitment to EPO.}
\begin{tabular}{l l l}
\specialrule{.2em}{.1em}{.1em} 
\multicolumn{3}{l}{\textbf{\begin{tabular}[c]{@{}l@{}}(a) How long do you spend (on average) \\ per week on EPO?\end{tabular}}}                                                                             \\ \specialrule{.2em}{.1em}{.1em} 
\textbf{Time Spent}                                                   & \textbf{\begin{tabular}[c]{@{}l@{}}On\\ Preparation\end{tabular}} & \textbf{\begin{tabular}[c]{@{}l@{}}On \\ Engagement\end{tabular}} \\ \specialrule{.2em}{.1em}{.1em} 
\begin{tabular}[c]{@{}l@{}}I don’t participate\\  in EPO\end{tabular} & 13 (10\%)                                                         & 13 (10\%)                                                         \\ \hline
0-1 Hrs                                                               & 77 (59\%)                                                         & 69 (53\%)                                                         \\ \hline
1-3 Hrs                                                               & 30 (23\%)                                                         & 33 (25\%)                                                         \\ \hline
3-5 Hrs                                                               & 7 (5\%)                                                           & 9 (7\%)                                                           \\ \hline
5-10 Hrs                                                              & 3 (2\%)                                                           & 5 (4\%)                                                           \\ \hline
\textgreater 10 Hrs                                                   & 1 (1\%)                                                           & 2 (2\%)                                                           \\ \specialrule{.2em}{.1em}{.1em} 
\end{tabular} \,\,\,\,\,\,\,\,\,\,
\begin{tabular}{l l l}
\specialrule{.2em}{.1em}{.1em} 
\multicolumn{3}{l}{\textbf{\begin{tabular}[c]{@{}l@{}}(b) How long would you like to spend (on average) \\ per week on EPO?\end{tabular}}}                                                                             \\ \specialrule{.2em}{.1em}{.1em} 
\textbf{Time Spent}                                                   & \textbf{\begin{tabular}[c]{@{}l@{}}On\\ Preparation\end{tabular}} & \textbf{\begin{tabular}[c]{@{}l@{}}On \\ Engagement\end{tabular}} \\ \specialrule{.2em}{.1em}{.1em} 
\begin{tabular}[c]{@{}l@{}}I don’t participate\\  in EPO\end{tabular} & 4 (3\%)                                                           & 4 (3\%)                                                           \\ \hline
0-1 Hrs                                                               & 60 (46\%)                                                         & 41 (31\%)                                                         \\ \hline
1-3 Hrs                                                               & 45 (34\%)                                                         & 55 (42\%)                                                         \\ \hline
3-5 Hrs                                                               & 13 (10\%)                                                         & 15 (12\%)                                                         \\ \hline
5-10 Hrs                                                              & 5 (3\%)                                                           & 10 (8\%)                                                          \\ \hline
\textgreater 10 Hrs                                                   & 4 (3\%)                                                           & 6 (5\%)                                                           \\ \specialrule{.2em}{.1em}{.1em} 
\end{tabular}
\label{tab:time}
\end{table}
}

We find a similar trend when comparing the actual and desired combined amount of preparation and engagement time. Using the intervals provided in the survey, we created four distinct bins of engagement and preparation, where low preparation (engagement) is defined as three hours or less and high preparation (engagement) is defined as more than three hours; i.e., a respondent in the low engagement-low preparation category could spend anywhere from 0-6 hours on preparation and engagement combined per week. When considering current practices, 86\% percent of respondents fall into the low engagement-low preparation group. Twelve percent ($n = 113$) of these respondents would like to increase their involvement, almost exclusively to the high engagement-high preparation group. The desire to move out of the low engagement-low preparation group is strongest amongst the youngest (age 18-30) researchers and lowest amongst the oldest (age 41+) researchers (16\% and 8\%, respectively). However, the overall desire to move to the high engagement-high preparation group is a consistent trend across age groups.

We also asked respondents to identify when they primarily engage in EPO (Q8). The majority indicated \textsc{I engage in EPO during work hours and during my free time} (63\%). The distribution of the remaining responses was as follows: \textsc{I only engage in EPO during my free time (i.e., during evenings and on weekends} (21\%), \textsc{I do not engage in EPO} (8\%), and \textsc{I only engage in EPO during work hours} (7\%). Two percent of respondents chose \textsc{I Don’t Know}. We analyzed this question by respondent demography and divided responses into two categories: those who engage in EPO only in their free time, and those who engage in EPO at work. This grouping was chosen such that we would have sufficient statistics for a chi-squared test. The group of respondents who engage in EPO during work hours and during free time was collapsed with the group who engage in EPO during work hours only. Our hypothesis here was that younger, more junior researchers would feel obligated to only engage in EPO during their free time rather than during work. Chi-squared tests show that differences between most demographic groups are not significant. However, we find some evidence that a larger fraction of temporary (26\%) versus permanent (6\%) respondents engage in EPO only during their free time ($\chi^{2} = 9.02, p = 0.0027$), which supports our hypothesis.

\subsection{Motives and Deterrents}
\label{subsec:mod}

Several survey items were designed to explore Research Questions 2 and 3 regarding why respondents may or may not engage in EPO activities.  

Respondents were asked if they feel responsibility toward engaging in EPO and how funding might impact their engagement. We hypothesized that: 1) more respondents would identify EPO engagement as a personal rather than professional responsibility, and 2) the proportion of respondents who indicate EPO should be a part of their professional responsibility would be larger than that who indicate EPO currently is a part of their professional responsibility. When asked whether they think engaging in EPO is part of their professional responsibility as a scientist (Q3), 69\% of respondents selected \textsc{Yes}. When asked if it should be part of their professional responsibility (Q4), this fraction rose to 76\%. When asked instead whether they believed it should be a personal responsibility of a scientist (Q5), 80\% responded \textsc{Yes}. Some respondents instead answered these questions with a conditional; in each case, less than 12\% responded \textsc{Yes, but only education (i.e., undergraduate teaching or mentorship)} and less than 4\% responded \textsc{Yes, but only public outreach (i.e., public lectures or volunteering at science festivals)}. We further examined these results by comparing the responses regarding perceived responsibility across the different demographic groups outlined above in Respondents. After performing chi-squared tests of independence we find that the differences between the fraction who selected \textsc{Yes} among these groups are not statistically significant ($p > 0.05$ for all comparisons). When asked how funding impacts EPO engagement, 17\% of respondents indicated they are currently funded specifically to participate in EPO, and 21\% indicated they hope engaging in EPO will help them secure future funding.

\begin{figure}[htp]
    \centering
    \includegraphics[width=\textwidth]{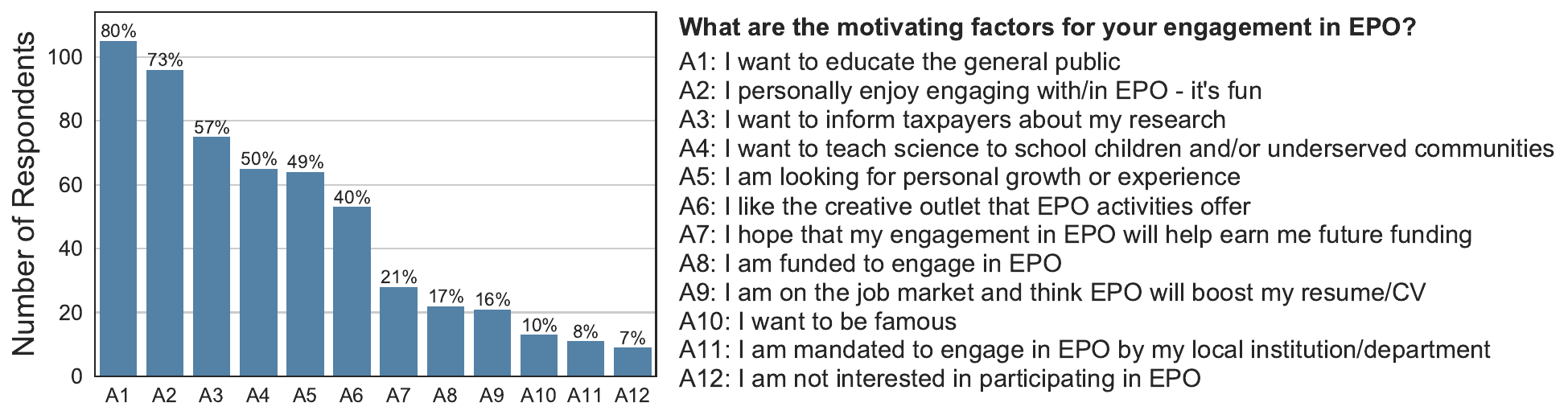}
    \caption{Distribution of checked motivating factors toward EPO engagement. Respondents were provided a list of possible motivating factors for EPO engagement and asked to check all that apply. }
    \label{fig:Fig3}
\end{figure}

Respondents were also asked about general motivations for engaging in EPO and any factors which deter their engagement. Figure~\ref{fig:Fig3} presents the distribution of responses for motivating factors (Q15). The most popular motivating factor for participating in EPO is the desire to educate the general public (80\%); this is closely followed by respondents engaging in EPO because they find it personally enjoyable (73\%). When asked about barriers to engagement (Q16), lack of time was overwhelmingly the most popular response (52\%). We note that in this survey item, there was no distinction made between time spent at work or personal time, or any conflict between spending time on EPO and research. Funding was also indicated to be an issue, as 19\% of respondents indicated they ``are not funded to do EPO.'' Additionally, 16\% of respondents indicated that they felt they lacked the skills and/or training to engage in EPO activities.

\begin{figure}[htp]
    \centering
    \includegraphics[width=\textwidth]{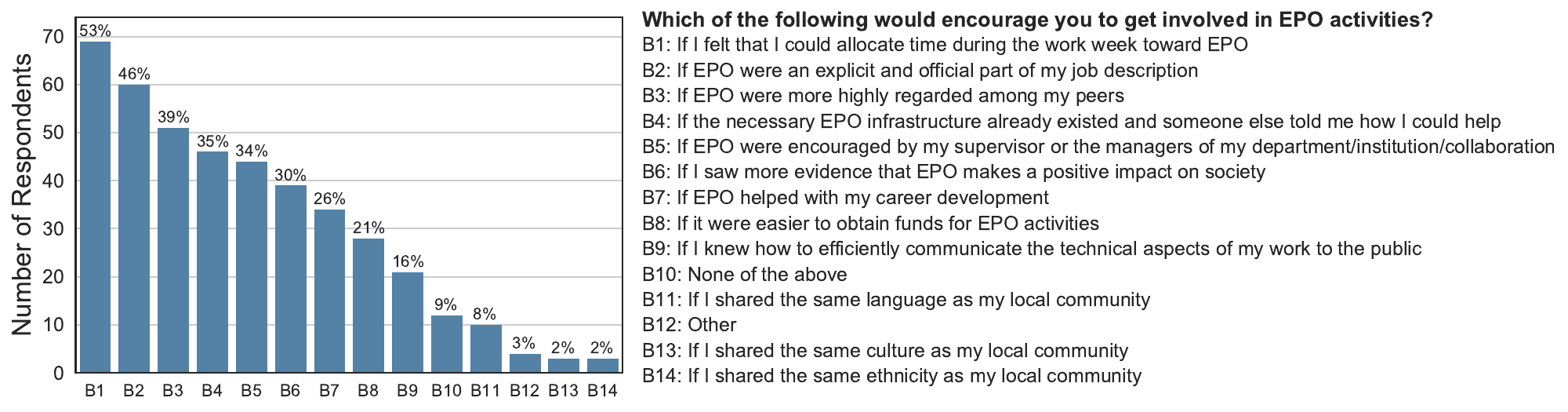}
    \caption{Distribution of checked factors which would encourage future EPO engagement. Respondents were provided with a list of possible incentives to encourage participation in EPO activities and asked to check all which might increase their motivation. Respondents could also write in their own responses; these have been combined into the ``Other'' category. 
}
    \label{fig:Fig4}
\end{figure}

We concluded this portion of the survey by asking researchers about factors which might encourage increased participation in EPO activities (Q17). Results are displayed in Figure~\ref{fig:Fig4}. The most frequent response was related to the previously discussed barrier of lack of time: 53\% of respondents noted that they would be more inclined to participate in EPO activities if they could allocate more time during the work week. Forty-six percent of respondents indicated that they would be more inclined to engage if EPO were listed as an explicit component of their job descriptions, and 39\% responded that their participation would perhaps increase if EPO were more highly regarded among their peers. 

To further explore the impact of these barriers, responses to Q17 were grouped into five barrier categories, Career Barriers (B1, B2, B5, B7), Academic Community Barriers (B3, B5), Impact Barriers (B6), Local Community Barriers (B11, B13, B14), and Skills Barriers (B4, B9), and then correlated with time spent on EPO engagement (Q9). A respondent was grouped into one of the five barrier categories if he identified at least one of the barriers in the categories in his response. When considering all responses, we find no significant correlations between time spent engaging in EPO and identification of each of the five barriers. We do find a trend that the strength of the relationship between identification of a Career Barrier and engagement time decreases with age; i.e., the youngest researchers are most strongly affected by Career Barriers. We also find that for the youngest researchers (age 18-31), there is a significant correlation between engagement time and identification of a Skills Barrier ($r = -0.40, p < 0.01$). The engagement of the mid-age researchers (age 31-40) is significantly correlated with identification of both the Academic Community Barrier ($r = -0.32, p = 0.06$) and the Local Community Barrier ($r = -0.33, p = 0.05$). 

\subsection{Centralization}
\label{subsec:centralization}
As a primarily grass-roots effort, analysis of the development and implementation of the DES EPO program \citep{Wolf:2018} offers important insight for future large collaborations. The final section of our survey focused on EPO in large collaborations, both for the purpose of self-reflection for the DES EPO organizational team, and to offer suggestions for future EPO programs (see Research Question 4). Specifically, we focused on understanding collaboration-members’ perception of EPO organized on a collaboration-wide scale, both in terms of administration and implementation. We refer to this collaboration-sponsored organization as a ``centralized'' EPO effort.

\subsubsection{Views of EPO Across the DES Collaboration}
\label{subsubsec:desviews}
Survey respondents were asked to rank the value they believe four DES-related groups place upon EPO (Q18). Respondents were asked to provide an answer using a 5-point Likert scale: 1 = \textsc{Not at All} to 5 = \textsc{Very Much}. Respondents were also given an \textsc{I Don’t Know} option. Table~\ref{tab:descent} summarizes the responses for four DES-related groups. The groups chosen for this item were intended to span the scope of an individual researcher’s involvement with DES, from a collaboration-wide level (the DES collaboration as a whole and those in collaboration management positions) to more personal interactions with other DES members (within a scientific working group or local to an institution). 

{\renewcommand{\arraystretch}{1.4}
\begin{table}[t]
\caption{Responses for the Likert survey item: ``Rank how much you think the following DES groups value EPO.'' Respondents were asked to rank on a 5-point scale: 1 = {\textsc{ Very Little}} to 5 = {\textsc{ Very Much}}. Respondents were also given the opportunity to answer {\textsc{I Don't Know}}. Percentages listed in the table correspond to the fraction from the total responses using the Likert scale only.}
\begin{tabular}{ l c c c c c c }
\specialrule{.2em}{.1em}{.1em}
\multicolumn{7}{l}{\textbf{Rank how much you think the following DES groups value EPO.}}                                                                                                                                                                                                \\ \specialrule{.2em}{.1em}{.1em}
\multicolumn{1}{c}{\textbf{\begin{tabular}[c]{@{}c@{}}DES\\ Group\end{tabular}}} & \textbf{\begin{tabular}[c]{@{}c@{}}(Very Little)\\ 1\end{tabular}} & \textbf{2} & \textbf{3} & \textbf{4} & \textbf{\begin{tabular}[c]{@{}c@{}}(Very Much)\\ 5\end{tabular}} & \textbf{I Don’t Know} \\ \specialrule{.2em}{.1em}{.1em} 
\begin{tabular}[c]{@{}l@{}}DES\\ Collaboration\end{tabular}                        & 1 (1\%)                                                            & 10 (9\%)   & 40 (35\%)  & 36 (31\%)  & 29 (25\%)                                                        & 15                    \\ \hline
\begin{tabular}[c]{@{}l@{}}DES\\ Management\end{tabular}                           & 4 (4\%)                                                            & 10 (10\%)  & 38 (37\%)  & 29 (28\%)  & 22 (21\%)                                                        & 28                    \\ \hline
Your DES Working Group                                                             & 11 (10\%)                                                          & 28 (26\%)  & 29 (27\%)  & 21 (20\%)  & 18 (17\%)                                                        & 24                    \\ \hline
\begin{tabular}[c]{@{}l@{}}DES members at your\\ Institution\end{tabular}          & 5 (4\%)                                                            & 22 (19\%)  & 29 (25\%)  & 35 (30\%)  & 26 (22\%)                                                        & 14                    \\ \specialrule{.2em}{.1em}{.1em} 
\end{tabular}
\label{tab:descent}
\end{table}
}

Generally, respondents indicated that each of the four DES groups place moderate to high value upon EPO. Overall, respondents indicated that their individual institutions placed the highest value upon EPO. From the perspective of an EPO facilitator, examining the number of \textsc{I Don’t Know
}responses offers a cursory assessment of EPO dissemination practices. We hypothesize that if EPO-related information is effectively communicated on behalf of and through various groups within DES that respondents will not be inclined to select the \textsc{I Don’t Know} option. The number of \textsc{I Don’t Know} responses was highest when asked about the value DES management places on EPO (21\%). The number of \textsc{I Don’t Know} responses related to the value DES working groups place upon EPO was similar (18\%). Respondents were better able to quantify the value the collaboration as a whole and individual member institutions place upon EPO; \textsc{I Don’t Know} responses for each of these groups was roughly 10\% of the total respondents. 

\subsubsection{Centralized Support for EPO Engagement}
\label{subsubsec:centralsupport}

Respondents were asked how DES and other large science collaborations could best support collaboration-wide engagement in EPO (Q22). The most frequent response (57\%) suggested that collaborations build and maintain a repository of talks, slides, curricula, etc., that can be used in various EPO activities (including both DES-sponsored and locally-organized programs). The second most frequent response involved funding: 54\% of respondents suggested collaborations could incentivize EPO participation by explicitly allocating funding for EPO projects. Another popular response (52\%) suggested that collaborations should consider EPO as time spent toward science infrastructure, which could lead to access to scientific data even after one leaves a collaboration member institution (i.e., data rights) or the ability to be a co-author on any collaboration publication (i.e., authorship rights: the DES Membership Policy document states that a member who spends a certain amount of time working on infrastructure can be granted data and authorship rights). Other popular responses included that collaborations hire dedicated EPO staff (48\%) and that collaborations could provide communication training for scientists (37\%).

In addition to checking predetermined answers for this survey item (Q22), respondents had the ability to write in responses. Of the seven written responses, four mentioned the role of collaborations in changing the cultural perspective of EPO within the physics community. Responses included calling for a ``change of community value'', instituting EPO engagement as an ``important factor in job applications'' for early career scientists, and calling for changing the perceived cultural norm that engaging in EPO is secondary, in terms of time and status, to research; e.g., ``making it acceptable to take time out of the work day to do EPO.'' Another point illuminated by the written responses was the desire to see a quantitative measure of the impact of EPO, by illustrating that the effort is ``impactful enough.''

Finally, researchers were asked to answer an open-ended question regarding the value of centralizing EPO efforts for large science collaborations (Q24). Eighty-one responses were coded by two researchers and organized into four categories: pro-centralization, anti-centralization, expressing confusion about the question and/or meaning of ``centralization,'' and Other. In some cases, a single response spanned multiple categories, e.g., a respondent lists both positive and negative aspects of a centralized EPO administration. Inter-rater reliability of these responses was 0.96. Roughly 84\% of respondents expressed ideas in favor of centralization. The most prominent response in favor highlighted that respondents believe centralization saves time and avoids duplication of effort. Twenty-eight percent of respondents presented ideas against centralization. The most common response against expressed concern that centralized efforts would replace more locally-organized events. Four percent of respondents expressed confusion about the meaning of ``centralized EPO.'' Overall, responses illuminated that respondents had differing views of the meaning of ``centralization.'' Some respondents interpreted it as an effort to facilitate EPO via making repositories or other means of coordination, while others had a more reductionist view, in which centralized EPO is a mechanism which replaces individual EPO activities. 

\section{Limitations and Future Work}
\label{sec:limitations}
Here we discuss some limitations of our survey and its analysis and list several possibilities for future studies. Firstly, we could have taken additional measures to improve survey validity and reduce response bias. While the survey was developed in collaboration with science communication professionals, we did not vet the survey using focus groups prior to distribution. Doing so would have helped bring attention to any confusing or ambiguous language or survey items. Also, before priming respondents with the \citet{Burns:2003} definition of EPO in the survey, we could have started by asking an open-ended question of how the respondents understand the term EPO. This might have helped us infer the respondents’ own familiarity with and attitudes toward EPO. Although, the first question of our survey is open-ended and asks the respondents to list the first three (or more) STEM EPO activities which come to mind, it is possible that question order may have biased responses. Coding and sorting the responses we do have could help determine what DES members consider to be EPO, though we leave this analysis for future work.

A second limitation of the survey is that questions were focused on how non-experts might learn from researchers, reinforcing the deficit model of engagement \citep{Besley:2011, Jensen:2016}. When listing options for what respondents think is the general purpose of EPO (Q2) or options for motivating factors for EPO engagement (Q15), there was no option that researchers can learn from the general public. Including this option would allow for a more balanced model of EPO as a conversation about science rather than a model of a public understanding of science. This would have expanded on the research of \citet{Martin-Sempere:2008}, who interviewed 167 science researchers who expressed that outreach should be understood as a dialogue between scientists and the public and who were motivated to engage in outreach by a desire to increase the public’s enthusiasm, awareness, and appreciation of science.

Thirdly, our findings regarding engagement barriers are intriguing. These responses could illuminate some particular areas of interest and discussion for those looking to facilitate EPO engagement. Responses to these questions could also potentially hint at the impact of the ``Sagan Effect,'' but we encourage future researchers to ask pointed questions about stigma more specifically and more directly. It could also be interesting to explore correlations between the various barriers to help illuminate if one barrier may be a proxy for another.

Finally, given that we surveyed only members of DES, it would be interesting to conduct additional studies with STEM researchers in domains outside of astrophysics and compare their responses to those presented in this article. Furthermore, many of our respondents were researchers at academic institutions; it would be useful to survey researchers working at museums or research institutes and see if they share similar beliefs.

\section{Summary and Conclusions}
\label{sec:summary}

In this article we present survey results from a programmatic study of the attitudes of astronomers, physicists, and astrophysicists toward EPO. The study was conducted using 131 responses from researchers in the international Dark Energy Survey collaboration. The survey was designed to explore general attitudes toward STEM EPO as well as those in the context of large-scale science collaborations. We note that as participation in the survey was voluntary, it is likely that respondents already had an interest in EPO engagement, resulting in a possible selection bias.

We find a disparity between the EPO activities in which respondents are involved (e.g., public presentations and teaching) and those that, in their opinion, would have more impact on the general public (e.g., on-air media and elementary or high school teacher development). We speculate that perhaps the respondents do not know how to personally effect change in the arena of formal education and that the opportunity to achieve such development via official organizations may be lacking. The low engagement we find in science writing and on-air media may be similarly explained due to their specialized and freelance nature — not many people have the skill or opportunity to perform such tasks. The true reason for these differences would be interesting to pursue in future studies.

Similar sentiments are reflected in the responses to questions of time commitment to EPO. Respondents reported that they spend less time preparing and engaging in EPO than they would like. This lack of time was also mentioned as the largest barrier preventing engagement in EPO, along with lack of funding, training and/or skills, and interest in performing organizational duties for EPO activities. As for what currently motivates EPO engagement, respondents stated a desire to educate the general public, reach minorities and underserved communities, and inform taxpayers of their work. Respondents also reported that they experience personal enjoyment from engaging in EPO, consider it as an opportunity for personal growth, and view it as a means to secure future funding (some government agencies require EPO components in their grant proposals). Furthermore, the majority of respondents believe that engaging in EPO-related activities is and should be a personal and professional responsibility of scientists.

In a comparison of when respondents engage in EPO, we find a significant difference in the behaviors of permanent versus temporary researchers. We hypothesize that this may be because temporary researchers feel that engaging in EPO during work hours is not appropriate or that their supervisors would not approve of such a use of work time. Respondents with permanent positions, however, may feel more in control of their time, or are perhaps even mandated by institutions and/or funding agencies to engage in EPO (particularly undergraduate education and mentorship) during work hours. This is further corroborated by the fact that a higher proportion of temporary researchers identify Career Barriers to EPO engagement than their colleagues with permanent positions. We believe that a similar analysis with a much larger sample size, as well as more carefully designed demographic groupings (e.g., for organizations wishing to increase EPO engagement amongst early career scientists), would be a compelling pursuit for a future study.

We also explored correlations between the amount of time researchers engage in EPO and identification of particular barriers to their engagement. Across all groups, we find a trend that identification of more career barriers correlates with less time engaging in EPO. For the younger researchers, we find a significant correlation between engagement time and barriers related to science-communication skills. This perhaps suggests that younger researchers feel they have not yet had the opportunity to develop skills they deem necessary for EPO engagement. We also find significant relations between engagement time and identification of barriers related to academic and local communities for researchers age 31-40. We hypothesize this may be due to the transient nature of the academic job search \citep{Petersen:2012}, and that researchers at this stage of their careers may not have yet had the opportunity to develop meaningful relationships in the workplace and/or with their local communities.

In addition to exploring EPO in the context of the individual researcher, we also sought to understand respondents’ attitudes toward EPO on a collaboration-wide level. When asked to rate how four different groups within DES value EPO, we find that although respondents generally indicate moderate to high value for each of the groups, the highest ranking is awarded to their individual member institutions (see Table~\ref{tab:descent}). We speculate that both the value assigned by respondents and the likelihood of not knowing what value to assign are determined primarily by how familiar (or not) the respondents are with the activities of the group in question. It follows, then, that DES members are more easily able to gauge the value placed on EPO by their home institutions; this would explain the high value assigned to this group and the low fraction of \textsc{I Don't Know} responses. It is also possible that since the primary goal of the science working groups is to engage in research, communicating scientific results outside of academia is not prioritized. Furthermore, it may be that the centralized EPO coordination did not extend throughout the hierarchy of DES infrastructure, i.e., there were no DES EPO representatives actively liaising between the EPO Committee and the working groups. Therefore, there was not an established channel of communication to regularly inform individual working-group members of EPO events. We believe these responses provide evidence for the need for EPO facilitators to think critically about how information regarding EPO-related activities is disseminated throughout the collaboration or organization. 

Responses highlight three key messages about centralizing EPO for collaborations. First, the responses suggest that effective collaboration-sponsored EPO programming, at least in the minds of collaboration researchers, requires a team dedicated to program organization, communication, and implementation. For example, building and maintaining a presentation slide and image repository is a substantial task which would require significant time and infrastructure expertise. Second, responses suggest the potential need for reevaluation of the allocation of EPO funding and the associated explicit directives for EPO engagement. This reconsideration of the funding stream is essential with respect to the collaboration leadership who are responsible for managing funds, but also possibly with respect to the greater sources of collaboration funding (i.e., government agencies and private foundations). Third, we find evidence that EPO leadership should clearly outline its role within a collaboration. Respondents who expressed concerns regarding ``centralized'' EPO efforts often remarked that such efforts would impinge upon time and effort toward locally-organized events. However, one of the goals of the DES EPOC was to provide support for these more individualized projects. \citet{Andrews:2005} found that the scientists in their study stressed the importance of having an outreach coordinator; in addition to being a source of EPO centralization, such a coordinator also helps break down barriers to engagement by saving the scientists time and providing them with specific information about potential outreach activities.
Based on the results presented here, we propose the following recommendations for those wishing to increase EPO participation amongst physicists, astronomers, and astrophysicists:
\begin{enumerate}
    \item \textbf{For scientists in positions of leadership at academic institutes or research facilities}
    \begin{enumerate}
        \item Facilitate opportunities for researchers to engage in public presentations and student mentoring, as they deem these activities most valuable.
        \item Develop a network for dissemination of scientific work through various media (e.g., online, press releases). 
        \item Foster institutional culture such that the time researchers spend engaging in and preparing for EPO activities most closely matches the time they \textit{would} like to be spending. 
        \item Create incentives like rewarding investment in EPO with benefits such as data rights and authorship on papers (Academic Community Barriers). Collaborate with EPO facilitators and funding agencies to develop further incentives.
        \item Discuss explicit expectations concerning time spent engaging in EPO in scientific job descriptions and interviews, independent of interviewee status, i.e., for graduate students, post-docs, faculty, etc. (Career Barriers, Academic Community Barriers).
        \item Establish a rapport within the institution and between the institution and the local community (Academic Community Barriers, Local Community Barriers).
        Consider allocating resources to a dedicated staff to organize, develop, facilitate, and evaluate EPO activities (Skills Barrier).
        \item Foster discussion of and roles directly related to EPO at all levels of a researcher’s involvement at the institutional and/or collaboration-wide level (Academic Community Barriers).
    \end{enumerate}
    \item \textbf{For EPO organizers and facilitators}
    \begin{enumerate}
        \item Survey researchers to determine which EPO activities they deem most valuable, and thus on which they would like to spend the most time. If feasible and necessary, provide assistance for implementation (Skills Barrier). 	
        \item Develop programming with realistic time commitments: likely no more than three hours of engagement and three hours of preparation per week.
        \item Collaborate with scientists in leadership positions and funding agencies to develop incentives for EPO participation (Academic Community Barriers).
        \item Design programming that provides the opportunity for, but does not rely on, researchers to develop a relationship with their local communities (Local Community Barriers).
        \item Organize internal programs targeting the development of science communication skills (Skills Barriers).
        \item Identify the roles of EPO organizers and facilitators amongst researchers. Describe how centralized EPO efforts fit into the context of EPO with local institutions and communities, as well as within the greater project.
    \end{enumerate}
\end{enumerate}

\section*{Acknowledgements}

The authors are grateful to E.~Bechtol, L. Biron, S. Pasero, M. S. S. Gill, and R. Cawthon for their input while writing the article. The authors are grateful to A. K. Romer, B. Nord, and J. Horwitz for survey design feedback.

\section*{Funding}

Funding for the DES Projects has been provided by the U.S. Department of Energy, the U.S. National Science Foundation, the Ministry of Science and Education of Spain, 
the Science and Technology Facilities Council of the United Kingdom, the Higher Education Funding Council for England, the National Center for Supercomputing 
Applications at the University of Illinois at Urbana-Champaign, the Kavli Institute of Cosmological Physics at the University of Chicago, 
the Center for Cosmology and Astro-Particle Physics at the Ohio State University,
the Mitchell Institute for Fundamental Physics and Astronomy at Texas A\&M University, Financiadora de Estudos e Projetos, 
Funda{\c c}{\~a}o Carlos Chagas Filho de Amparo {\`a} Pesquisa do Estado do Rio de Janeiro, Conselho Nacional de Desenvolvimento Cient{\'i}fico e Tecnol{\'o}gico and 
the Minist{\'e}rio da Ci{\^e}ncia, Tecnologia e Inova{\c c}{\~a}o, the Deutsche Forschungsgemeinschaft and the Collaborating Institutions in the Dark Energy Survey. 

The Collaborating Institutions are Argonne National Laboratory, the University of California at Santa Cruz, the University of Cambridge, Centro de Investigaciones Energ{\'e}ticas, 
Medioambientales y Tecnol{\'o}gicas-Madrid, the University of Chicago, University College London, the DES-Brazil Consortium, the University of Edinburgh, 
the Eidgen{\"o}ssische Technische Hochschule (ETH) Z{\"u}rich, 
Fermi National Accelerator Laboratory, the University of Illinois at Urbana-Champaign, the Institut de Ci{\`e}ncies de l'Espai (IEEC/CSIC), 
the Institut de F{\'i}sica d'Altes Energies, Lawrence Berkeley National Laboratory, the Ludwig-Maximilians Universit{\"a}t M{\"u}nchen and the associated Excellence Cluster Universe, 
the University of Michigan, the National Optical Astronomy Observatory, the University of Nottingham, The Ohio State University, the University of Pennsylvania, the University of Portsmouth, 
SLAC National Accelerator Laboratory, Stanford University, the University of Sussex, Texas A\&M University, and the OzDES Membership Consortium.

AAP is supported by the Jet Propulsion Laboratory. Part of the research was carried out at the Jet Propulsion Laboratory, California Institute of Technology, under a contract with the National Aeronautics and Space Administration. AAP also acknowledges the support from the Astronomical Society of the Pacific.

\bibliographystyle{apacite}
\bibliography{ref}

\end{document}